\documentstyle[manuscript,aps]{revtex}
\newcommand{\beq}{\begin{equation}}
\newcommand{\eeq}{\end{equation}}
\newcommand{\bea}{\begin{eqnarray}}
\newcommand{\eea}{\end{eqnarray}}
\newcommand{\beqa}{\begin{eqnarray}}
\newcommand{\eeqa}{\end{eqnarray}}

\def \A {{\bf A}}
\def \O {{\bf O}}

\def \did {{\bf \Pi_+}}
\def \didnt {({\bf 1}-\did)}

\def \proj {{ \bf \Pi}_{x_A}}

\def \J {{\bf J}}
\def \H {{\bf H}}

\def \P {{\bf P}}
\def \la {\langle}
\def \ra {\rangle}
\def \up {\uparrow}
\def \down {\downarrow}
\def \T {{\bf T}_A}

\def \la {\langle}
\def \ra {\rangle}

\def \x {{\bf x}}
\def \y {{\bf y}}
\def \p {{\bf p}}
\def \px {{\bf P}_x}
\def \py {{\bf P}_y}
\def \up {\uparrow}
\def \down {\downarrow}
\def \G {{\bf G}}

\input epsf

\begin{document}
\pagenumbering{arabic}
\title{Time as an Observable}

   \author{ { 
J. Oppenheim$^{(a)}$,\footnote{\it jono@physics.ubc.ca}
B. Reznik$^{(b)}$,\footnote{\it reznik@t6-serv.lanl.gov}
and W. G. Unruh$^{(c)}$}\footnote{\it unruh@physics.ubc.ca \\
}  
{\ } \\
(a) {\it \small   Department of Physics and Astronomy,  University of British
Columbia,
6224 Agricultural Rd. Vancouver, B.C., Canada
V6T1Z1}\\
(b) {\it \small Theoretical Division, T-6, MS B288, 
Los Alamos National Laboratory, Los Alamos, NM, 87545}\\
(c) {\it \small   CIAR Gravity and Cosmology Program, Department of Physics
and Astronomy,  University of British
Columbia,
6224 Agricultural Rd. Vancouver, B.C., Canada
V6T1Z1}
}

\maketitle

\begin{abstract}
The role of time in quantum mechanics is discussed.  The differences between
ordinary observables and an observable which corresponds to the time of an event is examined.  In particular, the time-of-arrival of a particle to a 
fixed location is not an ordinary quantum mechanical observable.
While we can measure if the particle arrives, 
we argue that the time at which it arrives always has an
inherent ambiguity.  The minimum inaccuracy of time-of-arrival measurements
is given by $\delta t_a > 1/E_k$ where $E_k$ is the kinetic energy of the
particle.  The use of time-of-arrival operators, as well as
current operators, is examined critically
\footnote{Many of the ideas presented here were first published in
our previous work, Y. Aharonov, J. Oppenheim, S. Popescu, B. Reznik, and W.G Unruh, Phys. Rev. A {\bf 57} 4130, (1998) (e-print, quant-ph/9709031),
Copyright 1998 by The American Physical Society. 
}.
\end{abstract}
 \section{Introduction}
It is often stated that the time of an event is not a standard observable
in quantum mechanics.  Unlike other observables such as
the position, momentum and energy of a particle, time is not represented
by an operator, but by a parameter $t$.  While this is certainly true,
it is important to realize that the difference between time, and other
observables is not merely formal.

For example, if at time $t$ a particle is detected at location $X$, then we
can say with certainty that at the same time $t$, the particle was not at
any other location $X'$.  However, if we turn on a detector
located at position $x$, and detect a particle at time $T$, then it is
quite possible that this particle might also have been detected at any
number of other times $T'$.  One can also find that the particle never
arrives at the location $x$, or that it is always at $x$.

More importantly, measurements happen {\it at a certain time}.  One
measures the particle's position at time $t$.  Even a continuous measurement
at a particular location is a series of measurements at a certain time.
Each instant that the Geiger counter doesn't click, it is measuring the
fact that a particle has not entered it.  Furthermore, operators which
are used to measure the time-of-arrival \cite{top}\cite{rovelli} to the location $x$, are not
measured at $x$, but rather at an instant in time.
In quantum mechanics, measurements made at different times can
disturb each other, which can make measurements of the time of an
event problematic.

%In a field theory, measurements of the field
%are carried out at a particular space-time point, however, time
%still has a special role.

In this paper, we are chiefly concerned with the time-of-arrival.
ie. the time that a particle first arrives to a particular location $x_A$.
One could also consider the time of first occurrence of any event.
For example, one can ask at what time the operator $\A$ first yields
the eigenvalue $a_i$.  The two cases are very similar.  In Section II
we discuss the general concept of time-of-arrival measurements.  In
particular, we argue that the current does not give a probability
distribution for the time-of-arrival.

In Section \ref{strigger} we present a model detector which can always
record whether a particle is detected at a particular location.
ie. contrary to previous claims made by Allcock \cite{allcock1},
who argued that the time-of-arrival cannot be measured accurately
due to the impossibility of locally absorbing a particle instantaneously, 
we show that one can always locally absorb an incoming particle and record that
it has arrived.
However, in \ref{sclock} we find that if we couple the system to a clock in 
order to measure the time
at which the particle arrives, then the particle will be reflected
without us being able to record the time.
The basic reason is that, unlike a classical mechanical
clock, in quantum mechanics the uncertainty in the clock's 
energy grows when its accuracy improves \cite{salecker}.

As a result, we find a minimum inaccuracy in measuring the time-of-arrival 
given by 
\beq
\delta t_A > 1/E_k.
\label{limit}
\eeq 
In Section IV
%\ref{sgen} 
we argue that (\ref{limit})
is a general feature of quantum mechanics.
In Section V we prove that a
time-of-arrival operator cannot exist.  In Section VI we
argue that the time-of-arrival operator 
does not correspond to the continuous measurements discussed
in this paper.   We conclude in Section VII
with a discussion of the main results.

\section{Time-of-arrival measurements}
In standard quantum mechanics, the
probability that a particle is found at a given location
at time $t$
is given by $|\psi(x=X,t)|^2$.  If we know $\psi(x,0)$ for all
$x$ then the system is completely described and we can easily 
compute this probability.  Using the Schr{\" o}dinger equation
we can also compute $\psi(x,t)$ at any time $t$.  This probability
corresponds to results of a measurement of position at a particular
time $t$.  Quantum mechanics gives a well defined answer to the question, 
"where is the particle at time $t$?"  

However, it is also perfectly natural to ask 
"at what time is the particle at a certain location."  This question is
often posed in the laboratory.  Here, quantum mechanics does not
seem to provide an unambiguous answer.
  
At first sight it seems that the simplest approach for measuring
the time at which a particle is at a location $x$
is to consider a detection process in which the detector
is activated only at an instant, $t=T$,  on each of the
particles of an $N$ particle ensemble.
Another measurement is performed at $t=T'$ on another
ensemble, etc.
The probability to find the particle is simply
$|\psi(x,t=T)|^2$, and  $|\psi(x,T)|^2N = n_T$ is
the average number of particles found at position $x$ at $t=T$.
However, $|\psi(x,T)|^2$, does  not represent
a probability in time, since it is not normalized properly.

One might be tempted therefore, to consider
\beq
{|\psi(x,t)|^2\over \int|\psi(x,t')|^2 dt' } \label{eq:probdist}
\eeq
This normalization can only be done if one knows the state
$\psi(x,t)$ at all times $t$ (infinitely far in the past and future).
If one would select only the cases of successful detection
and filter out null cases, one might be able to argue that this
expression corresponds to a
"relative probability" of the "branches".
(In the many-worlds interpretation where all branches
exist this may have some formal significance.)
However, the  expression above certainly does not yield
the probability to detect the particle at time t.

One reason for this failure is that a particle may be at
a location $x$ at many different times $t$.  I may find that
a particle has probability 1 of being 
at $x$ at time $t_1$, however, I cannot conclude that it
wasn't at $x$ at other times. In contrast,
if I measure an observable $\A$, then at any given time, the system 
will be measured to have only one eigenvalue of $\A$.  

This leads us to consider the {\it arrival-time} of a particle,
since a particle can only arrive once to a particular location.
In order to measure the arrival time, one cannot use the measurement
procedure above, since one needs to detect the particle at time 
$t_A$, and also know that the particle was not there at any previous time.
In other words, one must continuously monitor the location $x_A$ in order
to find out when the particle arrives.  However, this continuous measurement
procedure has it's own difficulty, and also emphasizes the problem with 
the previous probability distribution.  Namely, that  the probability to find 
a particle at $t=T$ is generally
$not$ independent of
the probability to find the particle at some other time $t=T'$.
ie. if $\proj$ is the projector onto the position $x_A$, then
in the Heisenberg representation
\beq
[\proj (t),\proj (t')] \neq 0.
\eeq
Measurements made at different times do not commute and will
disturb one another.
Therefore the probability distribution given by (\ref{eq:probdist}),
although well defined, does not give a probability distribution
{\it in time} \cite{tmeas}. Similar problems plague the current operator.  One can attempt to
use the operator $\J_+$ \cite{jad} given by
\beq
\frac{\partial \J_+}{\partial x} =\frac{d{\bf \Pi}_+(t)}{dt}
\eeq
where $\did = \int^{\infty}_{x_A} |x\ra\la x| dx $
to give the probability that the particle arrives between $t$ and $t+dt$. 
However, because the various $\did(t)$ don't commute with each other,
this does not give a probability in time. 
%Interestingly enough however, we find in section \ref{szero} that 
%wavefunctions with zero current do not trigger a model detector.

One may therefore try to define the probability that a particle arrives at
a certain time $t_k$ as the probability that it isn't at the location
$x_A$ at any of the times $t_1, t_2, ... t_j$ (where $j<k$), times the probability that it is at $x_A$ at $t_k$.
For a particle originally localized to the left of $x_A$ one can show
\cite{mielnik} \cite{tmeas} that this probability is given by
\beq
P(t_k) = \la \psi_0 | A_k | \psi_0\ra
\eeq
where
%\beq
%A_k = \U^\dagger\didnt \U^\dagger \didnt ...\U^\dagger \did
%\U ... \didnt \U \didnt \U
%\eeq
%and the probability that a measurement hasn't occurred is 
%\beq
%P(\down, t_k) = \la \psi_0 | B_k | \psi_0\ra
%\eeq 
%with
%\beq
%B_k=\U^\dagger \didnt \U^\dagger \didnt ...\U^\dagger \didnt
%\U... \didnt \U \didnt \U \label{eq:nomeas}
%\eeq
%By acting the unitary operators on the projection
%operators we can write the $A_k$ or $B_k$ in the "Heisenberg
%representation."
%For example 
\beq
A_k=\didnt(t_1)...\didnt(t_{k-1})  \did(t_k)
\didnt(t_{k-1})...\didnt(t_1).
\eeq
The operators  $A_k$ are not related
by a unitary transformation to $A_0$.  Nor is $A_k$ a
projection operator.  One can think of this measurement procedure as being 
akin to a Geiger counter located at $x_A$ which clicks when a particle enters it. At each $t_j$, a measurement is made to determine whether the particle
has arrived, and by making $\Delta =t_{i+1}-t_i$ as small as we wish,
we can model a continuous time-of-arrival measurement.  However, these probabilities are not universal.
In this case, they apply only to the particular measurement scenario
under discussion.  In particular, the probability distribution
is sensitive to the frequency $\Delta$ at which $\did$ is measured.
Each measurement of $\did$ disturbs the system, and can introduce large
frequency components in the momentum distribution of the particle.
If $\Delta$ becomes too small, than the particle is reflected without
being detected, a phenomenon which is related to the Zeno paradox \cite{zeno}.

\section{A Time-of-Arrival Measuring Device \label{smodel} }
\subsection{Triggering a Local Detector \label{strigger}}
 
The previous measurement scheme consisted of a series of measurements,
each of which collapses the wavefunction of the particle.  
As a result of the rapid sequence of measurements, the evolution of the
particle was disturbed.
One can
also consider models where the measurement is only made at some final time well after the particle has interacted with the detector.  Until this final
measurement, the evolution of the system is unitary.
In this section, we will consider a detector which includes a particle 
detector which switches the clock off as the particle arrives.
We shall describe the particle detector as a two-level
spin degree of freedom. 
The particle will flip the state of the trigger from 
"on" to "off", ie. from $\up_z$ to $\down_z$.
First let us consider a model for the trigger without
including the clock: 
\beq
H_{trigger} = {1\over 2m} {\bf P_x^2} + {\alpha\over2}(1 +
\sigma_x) \delta({\bf x}) .
\eeq
The particle interacts with the repulsive
Dirac delta function potential at $x=0$, only if the spin is in
the 
$|\up_x\ra$ state, or with a vanishing potential if the state is 
$|\down_x\ra$. In the limit 
$\alpha \to \infty$ the potential becomes totally reflective
(Alternatively, one could  have considered 
a barrier of height $\alpha^2$ and  width $1/\alpha$.)
In this limit, consider a state of an incoming particle and 
the trigger in the "on" state: $|\psi\ra |\up_z\ra$.
This state  evolves to 
\beq
|\psi\ra |\up_z\ra  \ \ \to \ \ 
 {1\over \sqrt2}
 \bigg[ |\psi_R\ra|\up_x\ra  + |\psi_T\ra  |\down_x\ra \biggr] ,\label{triggered}
\eeq
where $\psi_R$ and $\psi_T$ are the reflected and transmitted 
wave functions of the particle, respectively.

The latter equation can be rewritten as 
\beq
{1\over 2}|\up_z\ra (|\psi_R\ra +|\psi_T\ra)
+{1\over 2}|\down_z\ra(|\psi_R\ra - |\psi_T\ra) \label{reallytriggered}
\eeq

Since $\up_z$ denotes the "on" state of the trigger, and
$\down_z$
denotes the "off" state, we have flipped
the trigger from the "on" state to the "off" state with
probability $1/2$.
By increasing the number of detectors, 
this probability can be made as close as we like to one.
To see this, consider  $N$ spins as $N$ triggers 
and set the  Hamiltonian to be 
\beq
{\px}^2/2m + (\alpha/2)\Pi_n (1+\sigma_x^{(n)})\delta(\x) . 
\eeq
We will say that the particle has been detected if at least 
one of the spins has flipped.
One can verify that in this case 
the probability that at least one spin has flipped is now 
$1-2^{-N}$.

This model leads us to reject the arguments of Allcock.   He considers
a detector which is represented by a pure imaginary absorber $H_{int}=iV\theta(-\x)$. 
Allcock's claim is that measuring the time-of-arrival is equivalent to absorbing
a particle in a finite region. If you can absorb the particle in an arbitrarily short time, then you have succeeded in transferring the particle from an incident channel into a detector channel and the time-of-arrival can then be recorded. Using his interaction Hamiltonian one finds
that the particle is absorbed in a rate proportional to $V^{-1}$.  One can 
increases the rate of absorption by increasing $V$, but the particle will be reflected unless $V<<E_k$.  He therefore claims that since you cannot absorb the particle in an arbitrarily short time, you cannot record the time-of-arrival
with arbitrary accuracy.

However, our two level detector is equivalent to a detector which absorbs
a particle in an arbitrarily short period of time, and then transfers the information to another channel.  The particle is instantaneously
converted from one kind of particle (spin up), to another kind of particle
(spin down).  A model for arbitrarily fast absorption is also given in \cite{muga}, although in this case, the absorber does not work for arbitrary wavefunctions (it is momentum dependent).  
We therefore see that considerations of absorption alone do not place any restrictions on measuring the time-of-arrival.  However, as we shall
see in Section \ref{sclock}, adding a clock to the system will
produce a limitation on the accuracy of time-of-arrival measurements.

\subsection{Zero-Current Wavefunctions \label{szero}}

One interesting aspect of this detector, is that while it can be used for
wave-packets arriving from the left or the right, it will not always be 
triggered if the wavefunction is a coherent superposition of right and left
moving modes.  Consider for example, the superposition
\beq
\psi(x) = A e^{ikx} + A e^{-ikx}.
\eeq
One can easily verify that the current 
\beq
j(x,t) = -i\frac{1}{2m}\left[ \psi^*(x,t)\frac{\partial \psi(x,t)}{\partial x}
-\frac{\partial \psi^*(x,t)}{\partial x} \psi(x,t) \right]
\eeq
is zero in this case.  $[|\psi(0,t)|^2$ is non-zero, although the state
is not normalizable.
As in eq. (\ref{triggered}) this state evolves into
\beq
\la x|\psi\ra |\up_z\ra  \ \ \to \ \ 
 {A\over \sqrt2}
 \bigg[ (e^{ikx}+e^{-ikx})|\up_x\ra  + (e^{ikx}+e^{-ikx})  |\down_x\ra \biggr] 
\eeq 
Which, when rewritten in the $\sigma_z$ basis, is just
\beq
 A(e^{ikx}+e^{-ikx} ) |\up_z\ra .
\eeq
ie. the detector is never triggered.

This wavefuntion is similar to the antisymmetric wavefunctions discussed by
Yamada and Takagi in the context of decoherent histories \cite{yamada} and Leavens \cite{leavens} in the context of Bohmian mechanics, where also
one finds that the particles never arrive.  How to best treat these cases is an interesting open question.

\subsection{Coupling the Detector to a Clock \label{sclock}}

So far we have succeeded in recording the event of arrival to 
a point (modulo coherent antisymmetric wavefunctions). 
As of yet, we have no information at all on the time-of-arrival.
It is also worth noting that the net energy 
exchange between the trigger and the particle is zero, ie. the 
particle's energy is unchanged.

However, we shall see that when we proceed to couple the trigger to a clock
we do find a limitation on the time-of-arrival. The total 
Hamiltonian is now given by
\beq
H_{trigger+clock} = 
{1\over 2m} {\bf P_x^2} + {\alpha\over2}(1 + \sigma_x)
\delta({\bf x}) +  {1\over2}(1 + \sigma_z){\bf P}_y .
\eeq

The time-of-arrival is given by the variable ${\bf y}$ conjugate to $\P_y$. 
The accuracy of the clock $\delta t_A$ is given by $dy=1/dP_y$ so that
as the clock's accuracy increases, so does the coupling.
However, since we can have $\alpha>> P_y$ it would seem that 
the triggering mechanism need not be affected by the clock.
If the final wave function includes a non-vanishing  amplitude
of  $\down_z$, the clock will be  turned off
and the time-of-arrival recorded. 
However, the exact solution shows that this is not the case.
Consider for example an initial state
of an incoming wave from the left and the spin in the $\up_z$
state.

The eigenstates of the Hamiltonian in the basis of $\sigma_z$
are
\beq
\Psi_L(x) = \left( 
\matrix{
e^{ik_\up x}+\phi_{L\up}e^{-ik_\up x} \cr 
\phi_{L\down} e^{-ik_\down x}  \cr }
\right) e^{ipy} ,
\eeq
for $x<0$ and 
\beq
\Psi_R(x) = \left( 
\matrix{
\phi_{R\up} e^{ik_\up x} \cr 
\phi_{R\down} e^{ik_\down x}  \cr }
\right) e^{ipy} ,
\eeq
for $x>0$.
Here $k_\up = \sqrt{2m(E-p)}=\sqrt{2mE_k}$
and $k_\down =\sqrt {2m E} = \sqrt {2m(E_k+p})$.
 
Matching conditions at $x=0$ yields 
\beq
\phi_{R\up} = {{2k_\up\over m\alpha} - {k_\up\over k_\down}
\over {2k_\up\over m\alpha} - ( 1 +{k_\up\over k_\down}) }
\eeq
\beq
\phi_{R\down} =  {k_\up\over k_\down}((\phi_{R\up}-1)=
{  {k_\up\over k_\down}
\over {2k_\up\over m\alpha} - ( 1 +{k_\up\over k_\down}) } ,
\label{phidown}
\eeq
and 
\beq
\phi_{L\down} = \phi_{R\down} 
\label{leftdown}
\eeq
\beq
\phi_{L\up} = \phi_{R\up} - 1 .
\label{leftup}
\eeq

We find  that in the limit $\alpha \to \infty$ the transmitted 
amplitude is 
\beq
\phi_{R\down} = - \phi_{R\up} = {\sqrt{E_k} \over
\sqrt{E_k} + \sqrt{E_k+p} } .
\eeq
The transition  probability decays like $\sqrt{E_k/p}$.
From  eqs. (\ref{leftdown},\ref{leftup}) we get that
$\phi_{L\down}\to 0$, and $\phi_{L\up} \to 1$ as the accuracy of
the clock, and hence $p$, increases. As a result the 
particle is mostly reflected back and the spin remains in the 
$\up_z$ state; i.e., the clock remains in the "on" state.  
Without the clock, we can flip 
the "trigger" spin by means of a localized interaction, but when
we couple the particle to the clock, the probability 
to flip the  spin and turn the clock off decreases
gradually to zero when the clock's precision is improved.

Furthermore,  the probability distribution 
of the fraction which has been  detected depends on the accuracy 
$\delta t_A$ and can become distorted with increased accuracy.
This observation becomes apparent in the following simple example.
Consider an  initial  wave packet that is 
composed of a superposition of two Gaussians centered around $k=k_1$
 and 
$k=k_2>>k_1$. Let the classical time-of-arrival of the two Gaussians
be $t_1$ and $t_2$ respectively. When the inequality (\ref{limit})
is satisfied, two peaks around $t_1$ and $t_2$ will show up in 
the final probability distribution. 
On the other hand, for ${2m\over k_1^2}>
\delta t_A > {2m\over k_2^2}$, 
 the time-of-arrival of the less energetic peak will 
contribute  less to the distribution 
in $y$, because it is less likely to trigger the
clock.  Thus, the peak at $t_1$ will be suppressed.
Clearly,  when the precision is finer than $1/\bar E_k$
we shall obtain a distribution which is 
considerably different from that obtained for the case 
$\delta t_A > 1/\bar E_k$ when the two peaks contribute equally.

\section{General Considerations \label{sgen} }

In the above model, we found that the maximum precision for
measuring the time-of-arrival is given by
$\delta t_A > 1/E_k$.         

If the precision is made better than this, the particle is reflected.
Essentially, as Salecker and Wigner \cite{salecker} pointed out, the energy of the 
clock increases as its accuracy increases.  The particle, when it arrives,
must use its energy to turn off the clock, and if the clock's energy
is too large, then it is unable to do so.

First we should notice that this limitation does not seem to
follow
from the uncertainty principle. 
Unlike the uncertainty principle, whose origin is kinematic, 
this limitation follows from the nature of the  
$dynamic$ evolution of the system during a measurement.
Here we are considering a restriction on the measurement of a 
single quantity.

While the limitation only applies to the particular measurement model
discussed in the previous section,
there is reason to believe that it is a more general feature of quantum
mechanics.  

In the toy model considered above, 
 the clock and the particle
had to exchange energy $p_y \sim 1/\delta t_A$.
The final kinetic energy of the particle is larger by $p_y$.
As a result, the effective interaction by which the clock switches off, 
looks from the point of view of the particle like a 
step function potential. This led to  
``non-detection'' when (\ref{limit}) was violated.

Can we avoid this energy exchange between the 
particle and the clock? 
Let us try to deliver
this energy to some other system without modifying 
the energy of the particle. 
For example consider the following Hamiltonian for 
a clock with a reservoir:
\beq
H = {\px^2\over 2m} + \theta(-\x)H_c + H_{res} + V_{res}\theta(\x)
\eeq
The idea is that when the clock stops, it dumps 
its energy into the 
reservoir, which may include many other degrees of freedom, 
 instead of delivering it to the particle.
In this model, the particle is coupled directly 
to the clock and reservoir,
however we could as well use the 
idea used in the previous section.
In this case: 
\beq
H = {\px^2\over 2m} +{\alpha\over2}(1 + \sigma_x)
\delta({\bf x}) +  {1\over2}(1 + \sigma_z){H_c}+ 
 H_{res} + {1\over2}(1 - \sigma_z)V_{res} .
\eeq
The particle detector has the role of providing a coupling
between  the clock and reservoir.
 
Now we notice that in order to transfer the clock's
energy to the reservoir without affecting the free particle,  
we must also prepare the clock and reservoir 
in an initial state that satisfies the condition  
\beq
H_c - V_{res} =0
\eeq
However this condition does not commute with the clock time 
variable $\y$.  We can 
measure initially $\y - {\bf R}$, where $R$ is a collective 
degree of freedom of the reservoir such that $[{\bf R}, V_{res}]
= i$, 
but in this case we shall not gain information 
on the time-of-arrival $y$ since $R$ is unknown.
We therefore see that in the case
of a sharp transition, i.e. for a localized interaction with the 
particle, one cannot avoid a shift in the particle's energy.
The "non-triggering" (or reflection)
effect cannot be avoided.

In \cite{aharonov} we also examined 
a variety of detection models each of which yielded the limitation
(\ref{limit}). 
Many of these models, although simple, correspond to real experimental
procedures which are used everyday in the laboratory.  
For example, measurements usually 
involve some type of cascade 
effect, which lead to signal amplification and finally allows 
a macroscopic clock to be triggered. 
A typical example of this type would be the photo-multiplier where 
an initially small energy is amplified gradually and 
finally detected.  
Consider the following time-of-arrival detector
\beq
H = {\px^2/2m} + V(x)\py
\label{conti}
\eeq
where 
\beq
V(x) = \left\{ \begin{array}{ll}
-{x_A^2\over x^2} & x < x_A \\
-1  & x \ge x_A
\end{array}
\right. 
\eeq
Here $x_A$ is very small and positive.
As the particle rolls down the potential slope, its energy increases
and it is able to turn on the clock
\footnote{In this case, we can measure the time of arrival by subtracting
the clock time ${\bf y}$ from the time $t$ measured on another perfectly
accurate clock which is external to the system. }.
However, one can show that
the motion of the particle is affected, and one measures a disturbed
time-of-arrival.    
The basic problem with such a detector is that 
when (\ref{limit}) is violated, the ``back reaction''
of the detector on the particle, 
during the gradual detection, becomes large.
The relation between the final record to  
the quantity we wanted to measure is lost.

One can also imagine  introducing  a 
``pre-booster'' device just before the particle arrives 
at the clock. If it could   
boost the particle's kinetic  energy arbitrarily high, 
without distorting the incoming 
probability distribution (i.e. amplifying all wave components $k$
with the same probability),  
and at an arbitrary short distance from the clock, 
then the time-of-arrival could be measured to arbitrary accuracy.
Thus, an equivalent problem is: can we  
boost the energy of a  particle by 
using only localized (time independent) interactions?
  
In \cite{aharonov} we considered an energy booster
described by the Hamiltonian
\beq
H={1\over 2m} {\bf P_x^2} +\alpha\sigma_x\delta(\x) +\frac{W}{2}\theta(\x)(1+\sigma_z)
+\frac{1}{2}[V_1\theta(-\x)-V_2\theta(\x)](1-\sigma_z). 
\label{booster}
\eeq
and a particle incoming from the left initially in the $\up_z$ state.
Here, $\alpha, W, \ V_1$ and $V_2$ are positive constants.  $W$ damps out the 
$\up_z$  component of the wave function for $x>0$. $V_1$ damps out the  $\down_z$ component for $x<0$, and the $\down_z$ component has its energy 
boosted by an amount $V_2$ for $x>0$.

However, we were able to show that this fails in the general case. 
What happens 
is that while the detection rate increase, one generally destroys the 
initial information stored in the incoming wave packet. 
Thus although higher accuracy measurements are now possible, they do not 
reflect directly the time-of-arrival of the initial wave packet.

Finally we note, that while
it is difficult to provide a general proof for the case of time-of-arrival,
one can demonstrate in a model independent fashion,
that the inaccuracy relation (\ref{limit}) is necessary for measurements
of the traversal time \cite{traversal}. 
\section {Conditions on A Time-of-Arrival operator \label{scond} }
The time-of-arrival can be 
recorded by a clock situated at $x=x_A$ which switches off when the particle
reaches it.
%All that is needed in this measurement 
%is to switch the running clock off when
%the particle reaches the clock. 
In classical mechanics we could, in principle, achieve this 
with the smallest non-vanishing interaction between the particle
and the clock, and 
hence measure the time-of-arrival with arbitrary accuracy. 

In classical mechanics there is also another
indirect method to measure the time-of-arrival. 
First invert the equation of motion of the
particle and obtain the time in terms of the location 
and momentum, 
$T_A(x(t), p(t), x_A)$. 
This function can be determined
at {\it any time} $t$, either by a simultaneous
measurement of $x(t)$ and $p(t)$ and evaluation of $T_A$, 
or by a direct coupling to $T_A(x(t), p(t), x_A)$.
  
One drawback to this method, is that if one measures the function \\
$T_A(x(t), p(t), x_A)$ then one needs to know the full Hamiltonian 
for all time.  After
the measurement has occurred, one has to have faith that the 
Hamiltonian will not change after the measurement has been made. On the other hand, the continuous measurements we have described can be used with any Hamiltonian.

These two different methods, namely, the direct measurement,
and indirect measurement, are classically equivalent. 
They give rise to the same classical time-of-arrival.
They are not equivalent however,  in quantum mechanics 
%\cite{srinivas}.

In quantum mechanics 
the corresponding operator $\T ( \x(t), \p(t), x_A)$,
if well defined, can in principle 
be measured to any accuracy. On the other hand,
a direct measurement cannot determine the time-of-arrival
to greater accuracy that $1/E_k$

Still, one can imagine an indirect determination
of arrival time as described above,
by a measurement of some regularized time-of-arrival operator 
$\T (\x(t), \p(t), x_A)$ \cite{rovelli}.
An obvious requirement of $\T$ is that 
it is a constant of motion; i.e., the time-of-arrival 
cannot change in time.
As we shall show a Hermitian time-of-arrival
operator, with a continuous spectrum, 
can satisfy this requirement 
only for systems with an unbounded Hamiltonian.
This difficulty can however by circumvented by ``projecting out''
the singularity at $p=0$ and by using only measurements of $\T$
which do not cause a ``shift'' of the energy towards
the ground state.
Nevertheless, unlike the classical case, 
in quantum mechanics the result of 
such a measurement may have nothing to do with 
the time-of-arrival to $x=x_A$.

In the next two Sections  we shall  examine this 
operator and its relation to the continuous measurements described 
in the previous sections. 
First in this section we show that an exact time-of-arrival operator
cannot exist for systems with bounded Hamiltonian.  Allcock has
proven this for the simple case of a free particle \cite{allcock1}.

To begin with, let us start with the assumption that the 
time-of-arrival is described, as other observables in quantum 
mechanics, by a Hermitian operator $\T$. 
\beq
\T(t) |t_A\ra_t = t_A |t_A\ra_t
\label{eigen}
\eeq
Here the subscript $\ra_t$ denotes the time dependence of 
the eigenkets, and 
$\T$ may depend explicitly on time.
Hence for example, the probability distribution 
for the time-of-arrival for the state
\beq
|\psi\ra = \int g(t_A') |t_A'\ra dt_A'
\eeq
will be given by  $prob(t_A) = |g(t_A)|^2$. 
We shall now also assume that the spectrum of 
$\T$ is continuous and 
unbounded: $-\infty<t_A<\infty$.

Should $\T$ correspond to time-of-arrival it 
must satisfy the following obvious condition.
${\T}$ must be a  constant of motion and in the Heisenberg representation
\beq 
 {d{\T}\over d t}= 
 {\partial{\T}\over \partial t} +
{1\over i} [{\T}, H] = 0. 
\label{constant}
\eeq
That is, the time-of-arrival cannot change in time.  The fact that
the particle will (or did) arrive at 11 o'clock needs to be true at
all times.  If, at 9 o'clock, we find that the particle
will arrive at 11 o'clock, then if we make make the same measurement again
at 10 o'clock or at 12 o'clock, we should still find that the particle
will (or did) arrive at 11 o'clock.

For a time-independent Hamiltonian, time translation invariance
implies that the eigenkets  $|t_A\rangle_t$ depends only on $t-t_A$, 
i.e. the eigenkets  cannot depend on the absolute time $t$. 
This means for example that at the time of arrival: $|t_A\rangle_{t=t_A}=
|t_A'\rangle_{t=t_A'}$.
Time-translation invariance implies
%\footnote{In Allcock's proof \cite{allcock1} of the non-existence of a %time-of-arrival 
%operator for the special case of a free Hamiltonian, he assumes that
%$|t_a + \tau\rangle = e^{-i\tau H} |t_a\rangle$. However, by definition, a %time-of-arrival eigenstate which will arrive at the time $t_a$ will remain an
%eigenstate which arrives at $t_a$ as the system evolves.  Allcock's proof is 
%thus a proof 
%of the non-existence of a time operator -- not time-of-arrival operators.}
\beq
|t_A\rangle_t = e^{-i\G} |0\rangle_0 .
\eeq
where $\G= \G(t-t_A)$ is a hermitian operator.
Therefore,  $|t_A\rangle_t$ satisfies the differential equations
\beq
i {\partial\over \partial t_A} |t_A\rangle_t  = {\partial \G\over 
\partial t_A} |t_A\rangle_t= -{\partial \G\over 
\partial t} |t_A\rangle_t, \ \ \ \ 
i {\partial\over \partial t} |t_A\rangle_t  = {\partial \G\over 
\partial t} |t_A\rangle_t .
\eeq
Now act on  the eigenstate equation (\ref{eigen}) with the differential operators
$i\partial_{t_A}$ and $i\partial_t$. This yields
\beq
-\T {\partial \G\over  \partial t} |t_A\rangle_t = - t_A {\partial \G\over 
\partial t} |t_A\rangle_t + i|t_A\rangle_t , 
\eeq 
and 
\beq
i{\partial \T\over \partial t} |t_A\rangle_t  + \T {\partial \G\over 
\partial t} |t_A\rangle_t = 
t_A {\partial \G\over 
\partial t} |t_A\rangle_t . 
\eeq
By adding the two equations above, the dependence on $\partial \G/\partial t$ 
drops off,  and after using 
the constancy of $\T$ (eq. \ref{constant}) 
we get  
\beq
\biggl( [\T,\H] + i\biggr)|t_A\rangle = 0. 
\eeq
Since  the eigenkets $|t_A\rangle$ span,  by assumption,  the full Hilbert 
space   
\beq
[\T, \H] = -i . 
\label{conj}
\eeq
Hence $\T$ is a generator of energy translations. From equation (\ref{constant})
we have $\T= t - \hat {\bf T}$, where $\hat {\bf T}$ is the  
``time operator'' of the system whose Hamiltonian 
is $\H$.  It is well know that equation (\ref{conj}) 
is inconsistent unless the Hamiltonian is unbounded from above 
and below \cite{pauli}.
\section{Measuring the Time-of-Arrival Operator vs. Continuous Measurements}
Although formally there cannot exist a time-of-arrival operator $\T$, it
may be possible to approximate $\T$ to arbitrary accuracy.  
Kinematically, one expects that the time-of-arrival operator for a free particle arriving at the location $x_A=0$ might be given by 
\beq
\T=-\frac{m}{2}\frac{1}{\sqrt{\p}}\x (0)\frac{1}{\sqrt{\p }}. 
\eeq
In general, the choice for the time operator is clearly not unique due to
operator ordering difficulties. 
Furthermore, since $\T$  changes sign discontinuously at $p=0$, it's eigenvectors
\beq
\langle k | T \rangle = \left(  \theta(k)+i\theta(-k) \right)  \sqrt{\frac{k}{2\pi m}}
e^{i\frac{Tk^2}{2m}}   
\eeq
are not orthogonal.  
\beq
\langle T | T' \rangle = \delta(T-T') - \frac{i}{\pi(T-T')} .
\eeq
$\T$ is not self-adjoint.
We can however, define the regularized
Hermitian operator 
\beq
\T ' = O(\p)T\O(\p) 
\eeq
where $O(\p)$ is a function which is equal to $1$ at all values of $p$ except
around a small neighbourhood $\epsilon$.  For $|p|<\epsilon$, goes rapidly
to zero (at least as fast as $\sqrt{k}$).  
It's eigenvalues are
complete and orthogonal, and it circumvents the proof given above, 
because it satisfies  
\beq
[\T ', \H] = -i \O
\eeq
i.e. it is not conjugate to $\H$ around $p=0$.  Although $\T '$ is not
always the shift operator of the energy, the measurement can be carried
out in such a way that this will not be of consequence.  To see this,
consider the interaction Hamiltonian
\beq
\H_{meas} = \delta(t) {\bf q} \T ', 
\eeq
which modifies the initial wave function $\psi \to \exp(-iqT')\psi$.
We need to demand that $\T '$ acts as a shifts operator of the energy 
of $\psi$ during the measurement.  Therefore we need that 
$q>-E_{min}$,  where $E_{min}$ is the minimal energy
in the energy distribution of $\psi$. In this way, the measurement 
does not shift the energy down to $E=0$ where $\T '$ 
is no longer conjugate to $\H$.  The value of $\T '$ is recorded on 
the conjugate of $q$ -- call it $P_q$.  Now the uncertainty is given by
$dT_A' = d(P_q) = 1/dq$, thus naively from $dq=1/dT_A' < E_{min}$,  
we get $E_{min} dT'>1$.  However here, the average $ \la q \ra$ was 
taken to be zero.  There is no reason not to take $\la q \ra$ to be much 
larger than $E_{min}$,  so that $\la q \ra -dq >> -E_{min}$.  If we do so, 
the measurement increases the energy of $\psi$ and $\T '$ is always 
conjugate to $H$.  The limitation on the accuracy is in this case
$dT_A' > 1/ \la q \ra$ which can be made as small as we like.

However, even small deviations from the commutation relation
(\ref{conj}) are problematic.  Not only is the modification
arbitrary, it will also result in inaccurate measurements.
For example, since
\beq
\frac{d\T'}{dt}= {\bf 1} - \O,
\eeq 
\beq
\T '(t)=\T '(0) -t ({\bf 1}-\O).
\eeq
For the component of the wavefunction ${\tilde \psi(k)}$ which has support
in the neighbourhood of $k=0$, the time-of-arrival
will no longer be a constant of motion. The average value of $\T'(t)$
for the state ${\tilde \psi(k)}$ is given by
\beq
\la\T(t)\ra= \la\T(0)\ra -t\int dk \, \left[ 1-O(k) \right]
|{\tilde \psi(k)}|^2 \, .
\eeq
The second term on the right hand side will be non-zero if $\tilde{\psi(k)}$
has suppport for $|k|< \epsilon$.  
Even if ${\tilde\psi(k)}$ is negligibly small around $k=0$, the second term
will grow with time.  Thus, one only needs to wait a sufficiently long 
period of time before measuring $\T'$ to find that the average time-of-arrival
will change in time.  As mentioned in the previous section, this does not 
correspond to what one would want to call a "time-of-arrival".  The greater
$|{\tilde \psi(k)}|^2$ is around $k=0$, the greater the deviation from the condition that the
time-of-arrival be a constant of the motion.

Furthermore, careful examination of the eigenstates of the modified
time-of-arrival operator show that at the time-of-arrival, the
states have only a probability of $1/2$ of being found at the 
time-of-arrival \cite{normstates}.

%One also finds that at the time of arrival, the eigenstates of $\T$ or
%$\T'$  $\langle x | T(t_A)\rangle$ are not delta functions $\delta(x)$ but 
%are proportional to $x^{-3/2}$.  However, if one modifies 
%these eigenstates, by considering narrow superpositions of them given by

%\beq
%|\tau_\Delta\ra = \int  e^{-\frac{(T-\tau)^2}{\Delta^2}} |T\ra dT
%\eeq
%then, although the various $|\tau_\Delta\ra$ are no longer orthogonal,
%they are normalizable and by decreasing $\Delta$ one can
%make them as localized as one wishes around the point
%of arrival, at the time-of-arrival\cite{tsups}.  
%Also, at any time other than the
%time-of-arrival, one can make the probability that the particle is found
%at the point of arrival vanish as $\Delta$ goes to zero.

%Furthermore, if ${\bf P}_{0}$ is the projector onto $x=0$, one finds that
%\beq
%\langle\psi [\T, {\bf P}_{0}]|\psi\rangle=-\frac{i}{2}Re\{\psi(x=0)\int dk
%\psi^*(k)\frac{m}{k^2}\}. 
%\eeq  
%A measurement of the time-of-arrival operator is not equivalent to 
%continuously monitoring the point-of-arrival.  
%

Finally, how does the resulting measurement of a time-of-arrival operator
compare with that of a continuous measurement? 
From the discussion in Sections III and IV, it should be clear that in the limit
of high precision, continuous measurements respond very differently in comparison
to the time operator. At the limit of $dt_A\to 0$ all the particles
bounce back from the detector. Such a behavior does not occur for 
the time of arrival operator. Nevertheless, 
one may still hope that since the eigenstates 
of $\T$ have an  infinitely spread in energy, they do trigger
a clock   even if $dt_A\to0$.
For the type of models we have been considering,  
we can show however that this will not happen.

Let us assume that the interaction of one eigenstate of $\T$
with the clock  evolves as
\beq
|t_A\rangle|y=t_0\rangle \ \rightarrow   \ 
|\chi(t_A)\rangle|y=t_A\rangle  + |\chi'(t_A)\rangle|y=t\rangle.
\label{map}
\eeq
Here, $|y=t_0\rangle$ denotes an initial state of the clock with $dt_A\to0$, 
$|\chi(t_A)\rangle$ denotes the final state of the particle if the clock 
has stopped, and $|\chi'(t_A)\rangle$ the final state of the particle 
if the clock has not stopped.

Since the eigenstates of $\T$ form a 
complete set,  we can express any state of the 
particle as $|\psi\rangle= \int dt_A C(t_A) |t_A\rangle$.
We then obtain :
\beq
\int dt_A C(t_A) |t_A\rangle |y=t_0\rangle  \ \rightarrow  \ 
\int dt_A C(t_A) |\chi(t_A)\rangle |y=t_A\rangle +  \biggl( \int dt_A C(t_A) 
 |\chi'(t_A)\rangle \biggr) |y=t\rangle .
\eeq
The final probability to measure the time-of-arrival 
is hence $\int dt_a |C(t_a)\chi(t_a)|^2$. On the other hand 
we found that for a general wave function $\psi$, 
in the limit of  $dt_a\to 0$, the probability 
for detection vanishes.
Since the states of the clock, $|y=t_a\rangle$, 
 are orthogonal in this limit, this implies that  $\chi(t_a) = 0 $
in eq. (\ref{map}) for all $t_A$.
Therefore, the eigenstates of $\T$ cannot trigger the clock. 

It should be mentioned however, that one way of circumventing this difficulty may be to consider a coherent set of $\T$ eigenstates states instead of the 
eigenstates themselves. 
These normalizable states will no longer be orthogonal to each, so they
may trigger the clock if they have sufficient energy (although a wave packet which is a superposition of them with lower energy will not).  
In this regard it is interesting
to note that the average energy of a Gaussian distribution of time-of-
arrival eigenstates
is proportional to $1/\Delta$ where $\Delta$ is the spread of the Gaussian
\cite{normstates}.  
Since the probability of triggering
the clocks discussed in Sections III
%\ref{smodel} 
and IV
%\ref{sgen} 
decays as 
$\sqrt{E_k \delta t_A}$, the coherent states will not always trigger a clock
whose inaccuracy is $\delta t_A = \Delta$.
\section{Conclusion \label{sconc}}
We have argued that time plays a unique role in quantum mechanics,
and is unlike a standard quantum mechanical observable.
In the context of the time-of-arrival $t_A$,
we have found a basic limitation on 
the accuracy that $t_A$ can be determined reliably:
$\delta t_A > 1/\bar E_k$.
This limitation is quit different in origin 
from that due to the 
uncertainty principle; here it applies to the measurement of a $single$
quantity.
Furthermore, unlike the kinematic nature of the uncertainty
principle, in our case the limitation is essentially
dynamical in its origin; it arises when  the time-of-arrival is
measured 
by means of a continuous interaction between the 
measuring device and the particle. 

We have also argued that measuring whether the particle is at 
the location of arrival $x_A$ at various times, and also
measuring the current operator, do not allow one to construct
a probability distribution which one could interpret as representing
the probability that the particle will arrive at a certain time.

 We would also like to  stress that continuous measurements
differ both conceptually and quantitatively 
from a measurement  of the time-of-arrival operator.
Operationally one performs here two completely different measurements.
While the time-of-arrival operator is a formally
constructed operator which can be measured by an impulsive
von-Neumann interaction, it seems
that continuous measurements are much more closer to actual
 experimental set-ups.
Furthermore, we have seen that the result of these two measurements 
do not need to agree,   in particular  
in the high accuracy limit, continuous measurements 
give rise to entirely different behavior.
This suggests that as in the case of the problem of finding 
a ``time operator'' \cite{unruh-wald} for closed quantum systems, 
the time-of-arrival operator has a somewhat limited physical meaning.
 \\
\\
\\
{\bf Acknowledgments}
We would like to thank Arkadiusz Jadcyk and Philippe Blanchard
for organizing the symposium, and for inviting one of us 
to present our work.
\vfill \eject

\end{document}